\documentclass{sig-alternate}

\usepackage{subfigure}
\usepackage{algorithm}
\usepackage{algorithmic}
\usepackage{graphicx}
\usepackage{url}
\usepackage{color}

\begin{document}
%

\title{Adaptive Event Dissemination for\\Peer-to-Peer Multiplayer Online Games\footnotemark}

%
%
%
%
%

\numberofauthors{3} 
%
\author{
%
%
\alignauthor
Gabriele D'Angelo\\
\affaddr{Universit\`a di Bologna}\\
\affaddr{Mura A. Zamboni 7, I-40127 Bologna, Italy}\\
\email{g.dangelo@unibo.it}
\alignauthor
Stefano Ferretti\\
\affaddr{Universit\`a di Bologna}\\
\affaddr{Mura A. Zamboni 7, I-40127 Bologna, Italy}\\
\email{ferretti@cs.unibo.it}
\alignauthor
Moreno Marzolla\\
\affaddr{Universit\`a di Bologna}\\
\affaddr{Mura A. Zamboni 7, I-40127 Bologna, Italy}\\
\email{marzolla@cs.unibo.it}}

\date{\today}

\maketitle

\footnotetext{The publisher version of this paper is available at \url{http://dx.doi.org/10.4108/icst.simutools.2011.245539}. \textbf{{\color{red}Please cite as: Gabriele D'Angelo, Stefano Ferretti, Moreno Marzolla. Adaptive Event Dissemination for Peer-to-Peer Multiplayer Online Games. Proceedings of 2nd ICST/CREATE-NET Workshop on DIstributed SImulation and Online gaming (DISIO 2011). In conjunction with SIMUTools 2011. Barcelona, Spain, March 2011. ISBN 978-1-936968-00-8.}}}

\begin{abstract}
In this paper we show that gossip algorithms may be effectively used
to disseminate game events in Peer-to-Peer (P2P) Multiplayer Online
Games (MOGs). Game events are disseminated through an overlay
network. The proposed scheme exploits the typical behavior of players
to tune the data dissemination. In fact, it is well known that users
playing a MOG typically generate game events at a rate that can be
approximated using some (game dependent) probability
distribution. Hence, as soon as a given node experiences a reception
rate, for messages coming from a given peer, which is lower than
expected, it can send a stimulus to the neighbor that usually forwards
these messages, asking it to increase its dissemination
probability. Three variants of this approach will be
studied. According to the first one, upon reception of a stimulus from
a neighbor, a peer increases its dissemination probability towards
that node irrespectively from the sender. In the second protocol a
peer increases only the dissemination probability for a given sender
towards all its neighbors. Finally, the third protocol takes into
consideration both the sender and the neighbor in order to decide how
to increase the dissemination probability. We performed extensive
simulations to assess the efficacy of the proposed scheme, and based
on the simulation results we compare the different dissemination
protocols. The results confirm that adaptive gossip schemes are indeed
effective and deserve further investigation.
\end{abstract}




\section{Introduction}

Multiplayer Online Games (MOGs) are very demanding applications, that
require smart solutions to support the frequent interactions among
players occurring in a game session. Responsiveness and scalability
are probably the two main requirements that a distributed architecture
should provide in order to fully support MOGs. Scalable MOG
configurations can be obtained by replicating the game state at
different nodes on the network, so as to avoid the presence of a
single point of failure (i.e.~a single server). Different solutions
may be adopted. Among them, we cite mirrored server architectures,
where replicated servers are geographically distributed and clients
may connect to one of these
servers~\cite{Cronin:2004,Ferretti:2006:FGH,MullerFGM04}. Another
option is the use of Peer-to-Peer (P2P) architectures, where some kind
of overlay network is employed to distribute game events among game
participants~\cite{Ahmed:2008,debs,disio}. In this work, we focus on
P2P architectures.

As to responsiveness, smart schemes are needed for the dissemination
of game events among nodes participating to the same game session,
especially among those that maintain and manage a local version of the
game state. In fact, these nodes must quickly compute game
advancements, which are consistent with those computed by other nodes.

Previous works have already demonstrated that gossip strategies can be
proficiently employed to disseminate data in P2P overlay
networks~\cite{simutools,disio}. The topology of the overlay is a
critical aspect, since the different characteristics of the network
correspond to different features that influence the message
dissemination. For instance, when peers organize themselves as a
scale-free network, the network as a low diameter; hence, a low number
of hops is required to cover the whole network with a broadcast
message. However, certain nodes must act as hubs, i.e.~they maintain a
large number of neighbors (degree). This corresponds to an unbalanced
workload among peers.  Conversely, overlays with uniform degree
distribution result in balanced workload required to forward messages
in the network.

Regardless of the network structure employed as the P2P overlay, the
use of a dissemination strategy based on a static spanning tree built
on top of such overlay is not appropriate for highly dynamic
scenarios, such as P2P systems where the number of nodes (and hence
the network itself) frequently changes. At the same time, it is not
possible to employ pure broadcasts to disseminate game events. In
fact, the number of updates generated during a game session is quite
high. Hence, it is necessary to avoid as much as possible redundant
transmissions to not congestion the network.

The considerations above motivate the need to devise novel, adaptive
decentralized algorithms for game events distribution in dynamic P2P
systems. In this paper, we propose an adaptive gossip scheme that
exploits the typical behavior of game players to optimize the message
distribution among nodes. It is known that MOG players commonly
generate game events according to some (game-specific)
inter-generation probability distribution between successive
moves \cite{Armitage:2005}. This feature can be employed to
dynamically tune the dissemination probability of events coming from a
given peer in the overlay.

Specifically, based on our approach, messages are gossiped through an
overlay network (a mesh) using a completely decentralized
approach. Once a peer receives a message from a neighbor, it forwards
the message to other neighbors based on a dissemination probability.
As soon as a node $p$ observes that it is receiving messages from
another peer $q$ at a rate lower than expected, it activates a
countermeasure, asking its neighborhood (actually, the neighbor $n$
from which it usually receives messages originated from $q$ or a
random neighbor if it did not receive any message at all) to increase
its dissemination probability of game events. Three variants of the
this scheme are considered. According to the first one, upon reception
at $n$ of a request from $p$ to increase the event flow, $n$ increases
the probability of dissemination of game events towards $n$,
independently of the originator of the events to be forwarded. In the
second approach, $n$ increases its dissemination probability only for
game events originating from $q$ (independently from the
receivers). In the third variant, the node $n$ increases its
dissemination probability only for game events coming from $q$ and
that will be delivered to $p$, that is the specific sender that has
requested the probability increase.

The request from $p$ to $n$ to increase the dissemination probability
can be interpreted as a stimulus that remains active at $n$ for a
limited period of time. Then, the dissemination probability returns to
the original value (i.e.~the stimulus decades in time). This approach
is adopted to avoid that in time all dissemination probabilities reach
the maximum value and thus the gossip scheme becomes a pure broadcast
algorithm.

We assessed the algorithms above using simulation experiments. The
results demonstrates the effectiveness of our approach; in particular,
we we observe that adaptive gossip schemes improve the game event
dissemination with limited additional overhead. Additional
experimental and methodological aspects will be discussed in detail.

The reminder of this paper is structured as
follows. Section~\ref{sec:mod} presents the system model. In
Section~\ref{sec:alg} we discuss the adaptive gossip
algorithms. Section~\ref{sec:sim} reports on simulation experiments we
carried out to assess the viability of our proposal. Finally,
concluding remarks are reported in Section~\ref{sec:conc}.

\section{System Model}\label{sec:mod}

We consider a MOG system organized as a P2P network. Peers communicate
through an overlay, which means that only nodes which are directly
connected in the overlay can exchange messages. Nodes which are not
directly connected must resort to multi-hop communication. We do not
impose any restriction on the overlay, which can be generated using
any kind of algorithm and attachment protocol when peers join the
network. Several alternatives exist in the
literature~\cite{Barabasi2000,simutools,Fletcher04unstructuredpeer-to-peer}.

Each node produces game events which must be disseminated
to all other nodes in the network. 
The inter-generation time of
events follows a node-specific probability distribution.

It is clear that the topology of the overlay has a strong influence on
the performances of the content dissemination. For instance, if a
scale-free network is employed, then the network has a low diameter
(in general it ranges from $\log \log N$ to $\log N$, being $N$ the
number of nodes). This means that a message requires very few hops to
travel from a node to any other node, assuming that routing happens
only along shortest paths. Also, scale-free networks are known to be
resilient against random
failures~\cite{Newman03thestructure}. However, they contain a
non-negligible fraction of peers with high degree; these nodes (hubs)
have a high number of neighbors, and thus must maintain a high number
of active connections~\cite{Barabasi2000,simutools,guclu}. Hubs will
likely sustain a higher workload than the other low-degree nodes.

Conversely, if a network has uniform degree distribution, meaning that
all nodes have approximately the same number of neighbor nodes, then
the workload is equally shared among all peers. However, the diameter
of the network increases, and so does the number of hops needed to
cover the whole network with a broadcast~\cite{gridpeer}. Moreover,
random networks are more prone to partitioning after random
failures. Hence, some additional control mechanism is needed to cope
with this issue.

We assume that in the P2P system, every peer knows all other
peers. This is quite reasonable in a MOG, since all players may
interact with all other participants in the virtual world. In order to
reduce the amount of peers in the same network, several strategies may
be employed that, in essence, divide the virtual world in areas of
interest, hence restricting the interactions only among peers in the
same virtual region~\cite{Ahmed:2008,Iimura:2004,Yu:2005}. Messages
containing game events are distributed through the overlay. This
avoids that all peers must directly communicate with all other peers
in the network.

\section{Adaptive Gossip}\label{sec:alg}

The adaptive gossip algorithm\footnote{In the following of this paper,
  when referred to the proposed adaptive gossip schemes the terms
  ``algorithm'' and ``protocol'' will be used interchangeably.} is a
basic push scheme: nodes which have novel information to disseminate
are responsible for generating or forwarding messages to other
peers~\cite{disio}. Each node forwards messages (game events) to a random
subset of its neighbors. The peculiarity of our proposal is that the
dissemination probability of messages at node $p$ varies depending on
the communication performances perceived at $p$ during the game
session.

As mentioned, each peer $p$ knows the list of peers interacting in a given
area of interest of the virtual world. For each peer, $p$
maintains statistical information on received messages, such as the
average reception rate and the times of last received events. These
metrics allow $p$ to estimate whether it is receiving updates from
other peers at the ``correct'' rate. 

The gossip protocol is executed when $p$ receives a novel game event
generated at the application layer, or when a game event is received
from another peer $q$. In either case, $p$ randomly selects some
neighbors to which the event will be
propagated~\cite{conf/nca/GarbinatoRT07,verma}. In what follows, we
consider three different protocols that exploit a stimulus to increase
the probability of dissemination of messages at a given peer, based on
measured performances.

\subsection{Algorithm \#1: Stimuli Associated to Receivers}

The first protocol works as follows. When a node $p$ receives a new
message \emph{msg} from one of its neighbors $q$, then $p$ considers
all its neighbors (except $q$). For each neighbor $n$, the message is
forwarded to $n$ with probability $\upsilon_n$, where $\upsilon_n$ is
a threshold which is dynamically computed (details will be given
shortly). Note that each message is processed only once: we assume
that each node $p$ maintains a LRU cache of recently seen messages,
and drops all duplicates which are received after the first message
has been processed. The pseudo-code for this protocol is shown in
Algorithm~\ref{alg:gossip1}.

\begin{algorithm}[t]
\caption{Adaptive Gossip \#1: gossiping procedure executed by $p$}
\label{alg:gossip1}
\begin{small}
\begin{algorithmic}[1]
\REQUIRE{\emph{msg} generated at $p$ $\vee$ \emph{msg} received from a peer $q$}
\STATE $N_p \leftarrow p$'s neighbors $\setminus \ q$\hfill\COMMENT {$q =$ NULL if \emph{msg} originated at $p$}
\IF{\emph{msg} is a duplicate}
\STATE Return
\ENDIF
\FORALL{$\mathit{n} \in N_p$} %
    \STATE {currentTime $\leftarrow$ \textsc{getTime}()}
    \STATE $\upsilon_n \leftarrow$ \textsc{computeThreshold}($n$, currentTime) \label{line:calculateThr1}
    \IF{\textsc{random()} $< \upsilon_n$}  %
        \STATE \textsc{send}(\emph{msg},$\mathit{n}$)%
    \ENDIF%
\ENDFOR
\end{algorithmic}
\end{small}
\end{algorithm}

\begin{algorithm}[t]
\caption{Adaptive Gossip \#1: Monitoring procedure executed by $p$}\label{alg:mon}
\begin{small}
\begin{algorithmic}[1]
\LOOP
  \STATE \textsc{sleep}(monitoringPeriod)
  \STATE peerList $\leftarrow$ \textsc{retrievePeersLowRate}() \label{line:lowrate}\hfill\COMMENT{retrieve peers with low reception rate}
  \FORALL{$j \in$ peerList}
    \STATE $q \leftarrow$ \textsc{forwarder}($j$) 
\COMMENT{neighbor that sends msgs from $j$}
    \STATE \textsc{send}($q$, ``low rate from $j$'')
  \ENDFOR
\ENDLOOP
\end{algorithmic}
\end{small}
\end{algorithm}

\begin{algorithm}[t]
\caption{Adaptive Gossip \#1: Procedure executed by $p$ upon stimulus reception}\label{alg:reception}
\begin{small}
\begin{algorithmic}[1]
\REQUIRE stimulus received from $q$
\STATE \emph{timeLastStimulus}$_q$ = \textsc{getTime}()
\end{algorithmic}
\end{small}
\end{algorithm}

The threshold $\upsilon_n$ plays a fundamental role for the message
dissemination, since after such a gossip $p$ will never reconsider
$msg$ for dissemination.  All values $\upsilon_n$ are initially set to
a given constant $\upsilon_0$. Then, all values are updated
periodically by a monitoring procedure, which runs on each peer. The
monitoring procedure periodically measures the reception rate of game
events originated at each node $i$ in the overlay network. As soon as
$p$ observes a lower game event reception rate from a peer (say $j$),
it selects the neighbor $q$ from which it usually receives messages
containing game events generated by $j$;\footnote{When multiple peers
  are possible forwarders of messages originating from $j$,
$q$ is selected as a random neighbor of $p$.}  then, $p$
sends $q$ a stimulus message to request the increase of the value of
$\upsilon_p$ stored at $q$. This procedure is sketched in
Algorithm~\ref{alg:mon}.

The \textsc{retrievePeersLowRate}() procedure
(line~\ref{line:lowrate}, Algorithm~\ref{alg:mon}) collects the list
of nodes $i$, in the whole network, from which $R_i$ messages has been
received in the monitoring period $T_\mathit{mon}$ such that 
\[
R_i < \alpha \omega T_\mathit{mon}
\]
where $\omega$ is the expected event generation rate, and
$\alpha$ is a parameter that can be tuned.

The stimulus that increases the dissemination probability $\upsilon_p$
decays over time. This means that after a certain deadline, its effect
terminates and $\upsilon_p$ comes back to the default value
$\upsilon_0$.  Algorithm~\ref{alg:reception} describes what happens at
$q$ upon reception of a request from $p$ to increase its dissemination
probability: a variable \emph{timeLastStimulus}$_p$, which contains
the last received stimulus from $p$, is updated to the current
time. In fact, the measure of the threshold $\upsilon_p$ depends on
the time elapsed since \emph{timeLastStimulus}$_p$ as we will discuss
shortly.

\begin{figure}[t]
\centering\includegraphics[width=\columnwidth]{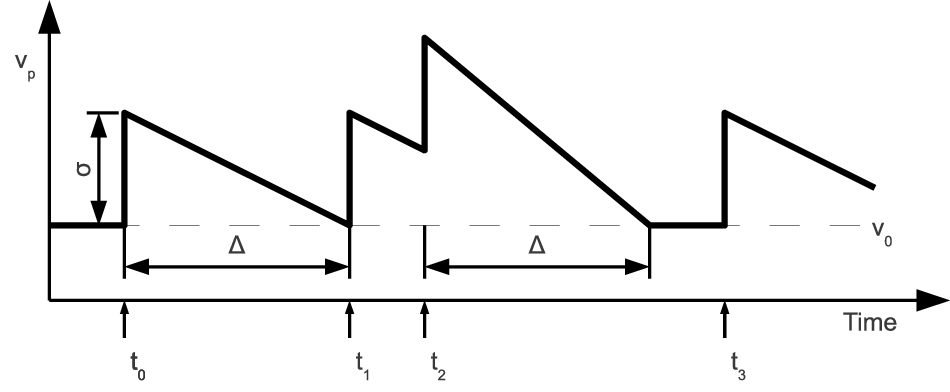}
\caption{Pictorial representation of function $\upsilon_p$. A stimulus
  $\sigma$ is received at times $t_0, t_1, t_2$ and $t_3$. At time
  $t_2$, the stimulus adds $\sigma$ to the current value of
  $\upsilon_p$. $\upsilon_p$ decays linearly to $\upsilon_0$ after
  time $\Delta$ from the last received stimulus.}\label{fig:stimulus}
\end{figure}

Different implementations of the \textsc{compute\-Thre\-shold}()
procedure (line~\ref{line:calculateThr1} in
Algorithm~\ref{alg:gossip1}) are possible. In the simulations, we
adopted the following function: all thresholds are initialized to a
default value $\upsilon_0$. Upon reception of a stimulus from a peer
$p$ at time \emph{timeLastStiumlus}$_p$, the actual value of
$\upsilon_p$ is increased by a fixed quantity $\sigma$. Then,
$\upsilon_p$ decades linearly over a time interval of length $\Delta$,
such that at time \emph{timeLastStimulus}$_p + \Delta$ its value is
back to $\upsilon_0$. If another stimulus is received during the
decaying phase, the stimulus adds to the current value of $\upsilon_p$
and \emph{timeLastStimulus}$_p$ is updated accordingly; in any case,
$\upsilon_p$ decays linearly to $\upsilon_0$ after time $\Delta$ from
\emph{timeLastStimulus}$_p$. We also remark that the value of
$\upsilon_p$ is limited to 1. See Figure~\ref{fig:stimulus} for a
pictorial explanation.

\subsection{Algorithm \#2: Stimuli Associated to Generators}

We now consider a protocol which differs from the previous one in the
method to adapt the dissemination threshold. Each peer $p$ maintains
an array of dissemination thresholds, one for each node in the
network. As soon as a new message \emph{msg} generated by $s$ has to
be disseminated by $p$, all $\mathit{p}$'s neighbors (except the peer
$q$ that sent \emph{msg}, if \emph{msg} has not been originated by $p$
itself) are considered and a threshold value $\gamma_s \leq 1$ is
computed.\footnote{Since the message is gossiped through an overlay,
  it is possible that $q \neq s$.} The value $\gamma_s$ is employed as
the threshold to determine if \emph{msg} has to be gossiped to a given
neighbor. The pseudo-code of the described protocol is outlined in
Algorithm~\ref{alg:gossip}.


\begin{algorithm}[t]
\caption{Adaptive Gossip \#2: the gossiping procedure executed at $p$}
\label{alg:gossip}
\begin{small}
\begin{algorithmic}[1]
\REQUIRE{\emph{msg} generated at $p$ $\vee$ \emph{msg} received from a peer $q$}
\IF{\emph{msg} is a duplicate}
    \STATE Return
\ENDIF
\STATE $N_p \leftarrow p$'s neighbors $\setminus \ q$\hfill\COMMENT {$q =$ NULL if \emph{msg} originated at $p$}
\STATE {$s \leftarrow$ peer that generated \emph{msg}}
\STATE {currentTime $\leftarrow$ \textsc{getTime}()}
\STATE $\gamma_s \leftarrow$ \textsc{computeProb}(s, currentTime) \label{line:calculateThr2}
\FORALL{$\mathit{n} \in N_p$} %
    \IF{\textsc{random()} $< \gamma_s$}  %
        \STATE \textsc{send}(\emph{msg},$\mathit{n}$)%
    \ENDIF%
\ENDFOR
\end{algorithmic}
\end{small}
\end{algorithm}

The monitoring procedure remains the same as the in previous scheme
(Algorithm~\ref{alg:mon}), as well as the function used to compute the
actual value of the threshold (\textsc{compute\-Thre\-shold}(),
line~\ref{line:calculateThr2} in Algorithm~\ref{alg:gossip}). The only
difference is that values $\gamma_s, \gamma_0$ must be employed
instead of $\upsilon_p, \upsilon_0$.  The same happens for what
concerns the request from $q$ to $p$ to increase the dissemination
probability related to peer $j$, i.e.~the procedure is equal, however,
the variable \emph{timeLastStimulus}$_j$ is employed to update the
variable $\gamma_s$.


\subsection{Algorithm \#3: Stimuli Associated to Generators and Receivers}

This version of the adaptive dissemination protocol is derived from
Algorithm \#2 and also in this case the difference is in the mechanism
used to adapt the threshold employed to gossip messages. In this
variant, each peer $p$ has not a single array of dissemination
thresholds (such as in Alg. \#2), rather it maintains a set of arrays
(one for each neighbor). In this way, each stimulus that is received
will cause a very selective update: it will change the probability to
disseminate the messages originated by a specific node which should be
forwarded to a given neighbor. The other parts of the dissemination
algorithm remain unaltered. The aim of this protocol is to generate
much more stimuli but each one is very specific and targeted.

\section{Simulation Assessment}\label{sec:sim}

In this Section we investigate the performance of the dissemination
protocols described above using a simulation mo\-del. First of all some
aspects about the comparison of dissemination protocols will be
discussed. Then the proposed algorithms will be compared to some
well known gossip protocols. Finally, some variants (i.e.~different
setups) of the protocol that gives the best outcomes will be shown.

\subsection{Testbed and metrics}\label{Testbed}

All dissemination protocols have been run on a set of $100$ different
overlay networks that have been randomly generated using an
Erdos-Renyi generator. All graphs are undirected, and they have been
constructed to ensure that they are also connected. Furthermore, we
ensure that the diameter of the graphs is always less than a
predefined value (i.e.~that will be used to set the Time-To-Live in
dissemination messages). As said before, also other graph structures
(i.e.~scale-free and small-world) would be interesting to evaluate but
are left as future work.\\

We now define some metrics under which the protocols will be
evaluated. Informally, a desirable property of a dissemination
protocol is that of being able to reach all nodes, and this should
happen as quickly as possible. Thus we define a metric called {\bf
  coverage}, which denotes the fraction of nodes which actually
received the messages. Ideally we wish to obtain $100\%$ coverage,
meaning that all nodes received all the generated messages. The second
metric is called {\bf delay}, and represents the average number of
hops that a message traverses before reaching a node (lower is
better). The delay is computed as follows: when a message is received
by a node for the first time, that node records the number of hops the
message traversed from its generation. The delay is computed as the
number of hops, averaged over all nodes which received the message,
and over all messages sent during a simulation run.\\

It is important to also define appropriate cost metrics, so that all
dissemination protocols can be compared in the same conditions. We define
the ``overhead ratio'' $\rho$ as follows:

\[
\rho = \frac{\textit{Delivered messages}}{\textit{Lower bound}}
\]

\noindent where ``delivered messages" is the total number of messages
that are delivered in a simulation run by a specific dissemination
protocol and the ``lower bound'' is the minimum number of messages (in
each graph) that are necessary to obtain a complete coverage. Thus,
the lower bound represents the number of messages sent by a broadcast
protocol which deliver events along the edges of a spanning tree, and
never sends duplicates. The lower bound depends on the graph and is
independent from the dissemination protocol to be used. For example,
in a graph of $n$ nodes and in which $m$ different events are
generated, the lower bound to the number of delivered messages is
$\Omega(n m)$. Each newly generated message has to traverse at least
$n-1$ links to eventually reach all nodes in the graph. Observe that
$n-1$ is precisely the number of edges on any spanning tree on a graph
with $n$ vertices.


\subsection{Simulator}

The performance evaluation of the adaptive gossip algorithms described
in Section~\ref{sec:alg} has been conducted using discrete-event
simulation based approach. The simulator, called LUNES (Large
Unstructured NEtwork Simulator)~\cite{pads} has been rewritten from
scratch after our previous work on PaScaS~\cite{simutools}.\\

The main goal of LUNES is to offer an efficient and easy-to-use tool for the
simulation of complex protocols on top of large graphs. In practice,
LUNES is able to import the graph topologies generated by other tools
(e.g.~igraph) and provides the functionalities that are needed for the
performance evaluation of simulated protocols. One of the main goals
of the simulator redesign is to obtain a tool that clearly splits the
fundamental phases:
\begin{itemize}
\item network topology creation;
\item protocol simulation in a specific testbed;
\item traces analysis (i.e.~performance evaluation).
\end{itemize}
This modular approach permits the easy integration of external
software tools. In practice, such integration is based on very simple
template files (such as the graphviz dot language~\cite{graphviz}) and
a provides a good level of extensibility. Under the performance and
scalability viewpoint, the most demanding points are the protocol
simulation and the traces analysis. The first one is demanded to a
specific simulator that will be discussed shortly. The second one,
that is the traces analysis, has been excluded from the simulation
tasks and some specific software tools have been implemented.

The amount of traces generated by a single run of the protocols
described in this paper, in a medium size graph, is pretty large. For
example, to evaluate the performance of a dissemination protocol all
the different messages seen by each node have to be accounted
for. Therefore, efficiency is essential in order to obtain timely
results. In the current version of LUNES this task is implemented via
a mix of shell scrips and dedicated tools written in C language (for
performance reasons). All such tools have been designed and
implemented to work in parallel and therefore are able to exploit all
the computational resources provided by parallel (multi-processor or
multi-core) architectures.\\

As said above, the simulation services are demanded to the ART\`IS
middleware and the GAIA framework~\cite{pads}. In this way, the LUNES
user does not need to deal with low-level simulation details and can
transparently take advantage of all the features offered by
GAIA/ART\`IS. In particular, to allow the simulation of large models,
a parallel and distribution simulation approach can be followed and
some advanced features such as the dynamic model partitioning and
load-balancing features are implemented transparently. For example,
clusters composed of very heterogeneous nodes (in terms of hardware)
can be employed for the simulation of large networks: the model
partitioning is dynamic, adaptive and totally demanded to the
simulation tools without any tuning to be done by the simulator
user~\cite{gda-ijspm-2009}.

\subsection{Model parameters}

In the following we have considered networks (i.e.~graphs) composed of
$100$ nodes (i.e.~peers) and generated as reported in
Section~\ref{Testbed}. Each node has $2$ edges, that is $200$ edges in
the whole network. Given the good scalability of the simulator, the
evaluation of graphs with a larger number of vertices and edges is not
a problem. This task is left as future work given that now we are more
interested in a preliminary validation of the proposed approach.

Focusing again on the model parameters, each simulation run is $5000$
time steps long and each node in the network can generate new messages
during the whole simulation lifespan. The time between successive
messages is generated according to a typical exponential distribution.
The variability of the inter-generation between two successive game
events generated by the same peer is of main importance, since peers
send stimuli to their neighbors based on the variability of reception
of messages. In other words, it is important to understand whether a
low game event reception rate is due to a poor dissemination caused by
the gossip algorithm, rather than a low generation rate by a given
node. It is worth noting that other ``more predictable'' distributions
(e.g.~uniform) would largely increase the results of all adaptive
dissemination algorithms. That's because each form of variability has
the effect to ``confuse'' the evaluation heuristics implemented in the
adaptive gossip protocols and to generate stimuli that are not
necessary, with the side effect to increase the communication cost of
the protocol.

Finally, as already introduced in Section~\ref{sec:alg}, each node
implements a cache structure with the aim to reduce the number of
duplicate messages sent around. This cache is managed using the Least
Recently Used (LRU) replacement algorithm; the cache size has been set
to $256$ items. To limit the lifetime of each message in the network,
we implemented a Time-To-Live (TTL) scheme. When a new message is
created, the TTL is set to 8, a value that we ensure is always greater
than or equal to the network diameter. As usual, each hop will reduce
this value up to discarding.

\subsection{Results}

We first compare the simplest adaptive algorithm (i.e.~\emph{Algorithm
  \#1: Stimuli Associated to Receivers}) with respect to very common
dissemination algorithms such as the \emph{Fixed Probability} and the
\emph{Probabilistic Broadcast} disseminations. In the \emph{Fixed
Probability} dissemination scheme (see Algorithm~\ref{alg:fixedprobability}), the node that receives a new message
randomly selects those edges through which the message must be
propagated. In the \emph{Probabilistic Broadcast} (Algorithm~\ref{alg:probabilisticbroadcast}), the node decides
whether to forward the received message with a certain probability. If
the message is forwarded, it is always sent to all neighbors. For more
details, see~\cite{disio,simutools}. We have already shown that
another very common dissemination algorithm, called \emph{Fixed Fanout}, is
unable to offer acceptable results in these conditions~\cite{disio}.

\begin{algorithm}[t]
\caption{Fixed Probability dissemination}
\label{alg:fixedprobability}
\begin{small}
\begin{algorithmic}[1]
\REQUIRE{\emph{msg} generated at $p$ $\vee$ \emph{msg} received from a peer $q$}
\STATE $N_p \leftarrow p$'s neighbors $\setminus \ q$\hfill\COMMENT {$q =$ NULL if \emph{msg} originated at $p$}
\IF{\emph{msg} is a duplicate} \STATE Return
\ENDIF
\FORALL{$\mathit{n} \in N_p$} %
    \IF{\textsc{random()} $< \upsilon_0$}  %
        \STATE \textsc{send}(\emph{msg},$\mathit{n}$)%
    \ENDIF%
\ENDFOR
\end{algorithmic}
\end{small}
\end{algorithm}

\begin{algorithm}[t]
\caption{Probabilistic Broadcast dissemination}
\label{alg:probabilisticbroadcast}
\begin{small}
\begin{algorithmic}[1]
\REQUIRE{\emph{msg} generated at $p$ $\vee$ \emph{msg} received from a peer $q$}
\STATE $N_p \leftarrow p$'s neighbors $\setminus \ q$ \hfill\COMMENT {$q =$ NULL if \emph{msg} originated at $p$}
\IF{\emph{msg} is a duplicate} \STATE Return
\ENDIF
\IF{\textsc{random()} $< \upsilon_0$ $\vee$ \emph{msg} generated at $p$}  %
\FORALL{$\mathit{n} \in N_p$} %
        \STATE \textsc{send}(\emph{msg},$\mathit{n}$)%
\ENDFOR
\ENDIF%
\end{algorithmic}
\end{small}
\end{algorithm}

In this first test, the setup of the adaptive algorithm is the
following: $monitoringPeriod=100$, $\sigma=0.2$, $\delta=300$,
$\alpha=1/3$ (see Section~\ref{sec:alg} for the description of each
parameter). For each algorithm, at least $10$ different setups have
been evaluated and the best results are shown.

\begin{figure}
\centering
\includegraphics[width=6.0cm,angle=270]{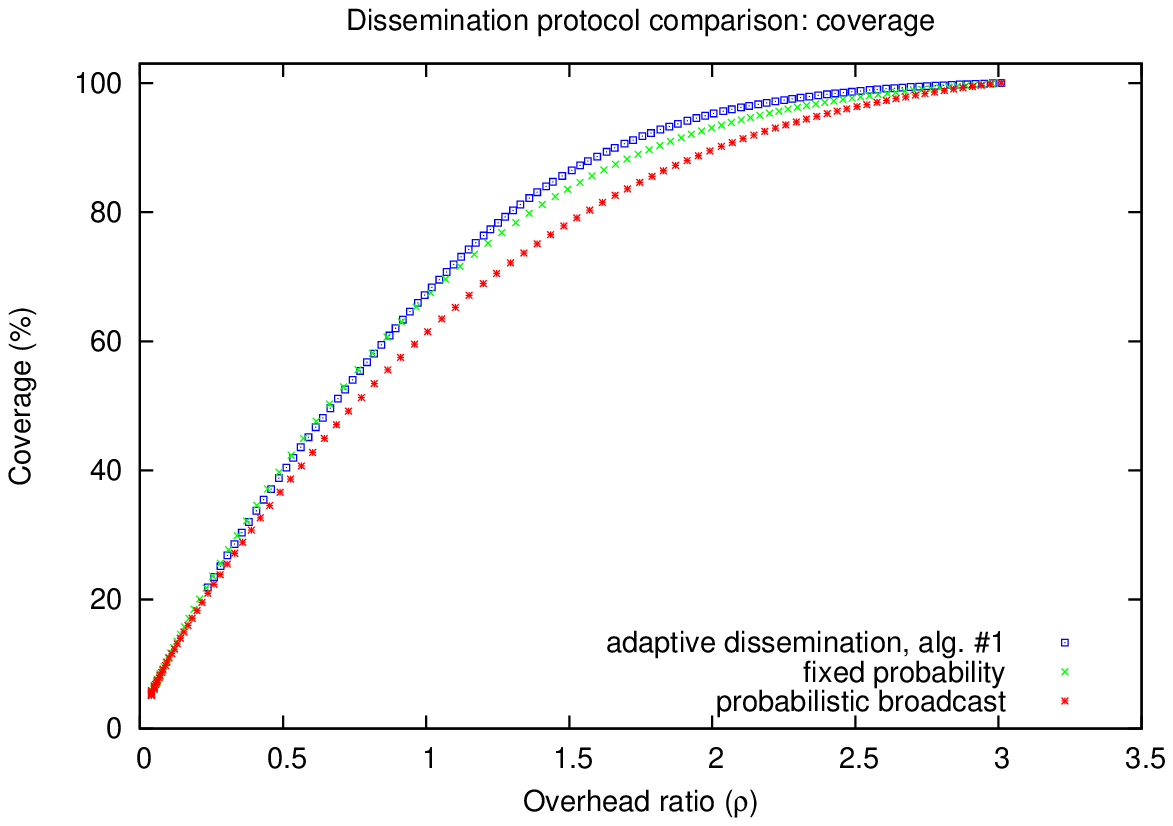}
\caption{Dissemination protocols, coverage}\label{fig:1_coverage}
\end{figure}


The Figures in this Section show the results obtained by the proposed dissemination protocols (in terms of {\bf coverage} and {\bf delay}) with respect to the cost (in terms of ``overhead'', $\rho$) incurred by each protocol. Given a dissemination protocol and a specific setup, we varied a single parameter: the default dissemination probability $\upsilon_0$. Specifically, we considered 100 different values for $\upsilon_0$, uniformly distributed in the range $(0,1]$. For each value of $\upsilon_o$, we executed each algorithm on multiple different random graphs, all of the same size, and computed average performance results on each set of runs. The resulting data set is shown in the Figure.

In Figure~\ref{fig:1_coverage} we observe that, in terms of {\bf
  coverage}, the adaptive gossip protocol (\emph{Alg. \#1}) is much
better than the \emph{Probabilistic Broadcast} and slightly better
than \emph{Fixed Probability}. The results obtained for $\rho<1$ are not very interesting given that, in any
case, all protocols would be unable to obtain a full coverage of the
network (because the total number of messages used for the
dissemination is below the lower bound seen above). On the other hand,
when $\rho>2.5$ the network is so full of messages that all the
algorithms have very similar outcomes.

\begin{figure}
\centering
\includegraphics[width=6.0cm,angle=270]{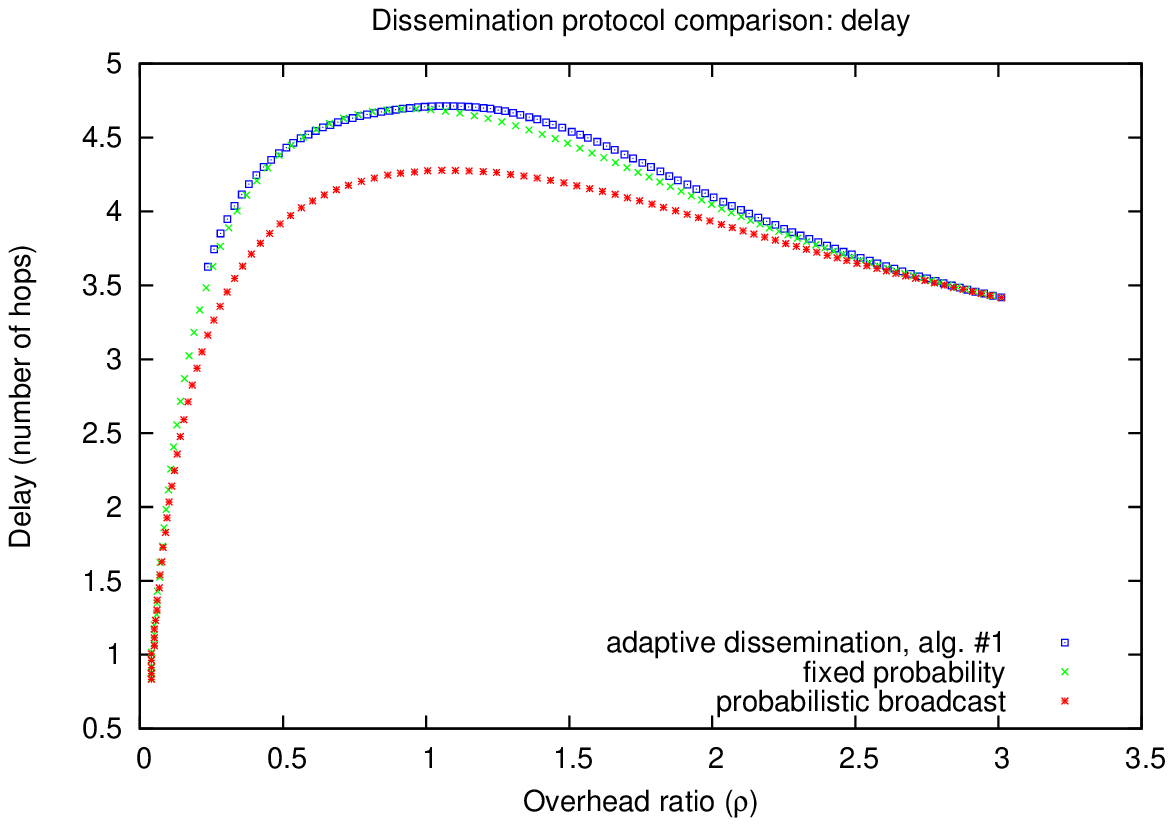}
\caption{Dissemination protocols, delay}\label{fig:1_delay}
\end{figure}

In terms of {\bf delay}, the \emph{adaptive protocol} is slightly
worse than the \emph{Fixed Probability}. It is worth noticing that, in
this case, lower is better given that the delay is proportional to the
average time that is necessary for the delivery of messages.

Given the results obtained above, in the following of this performance
evaluation, we will consider only three different flavors of the
adaptive gossip algorithms (as described in
Section~\ref{sec:alg}). Furthermore, for the reasons described few
lines above, results will be shown only for $\rho\geq1$.

\begin{figure}
\centering
\includegraphics[width=6.0cm,angle=270]{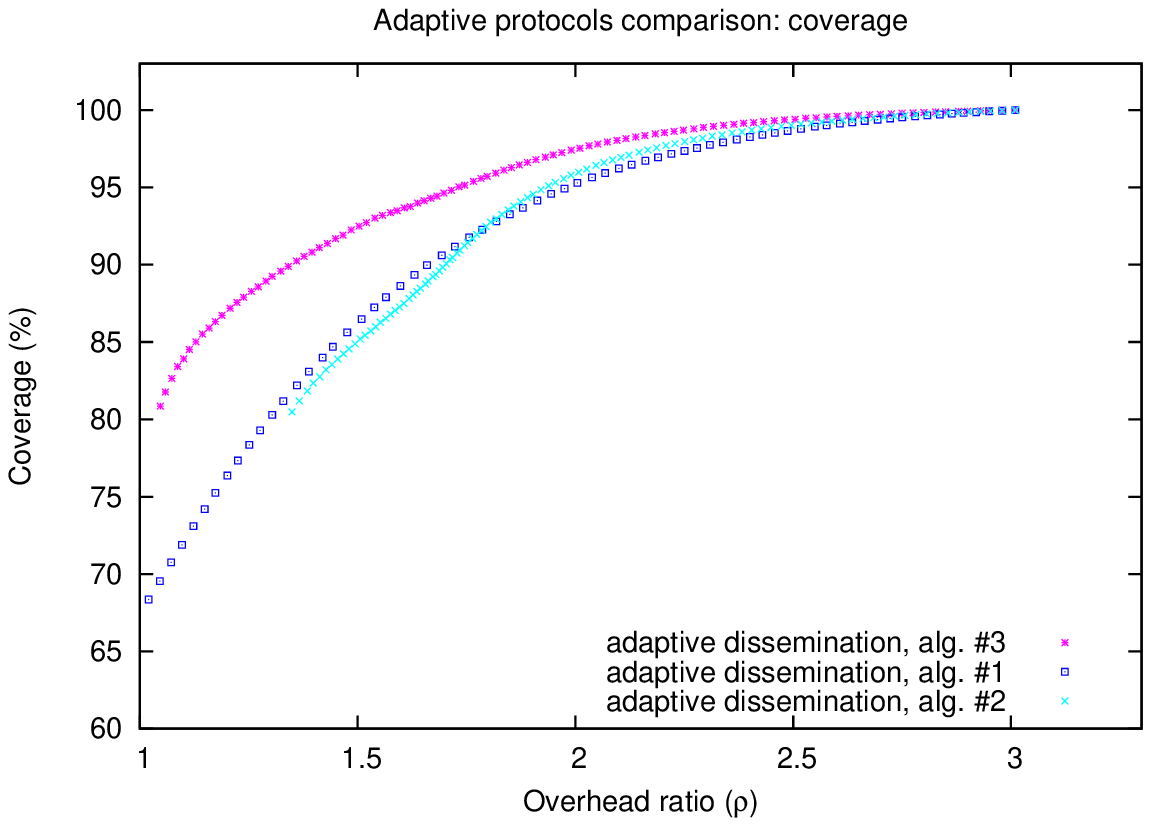}
\caption{Adaptive protocols, coverage}\label{fig:2_coverage}
\end{figure}

As expected the more complex adaptive algorithm (i.e.~\emph{Algorithm
  \#3: Stimuli Associated to Generators and Receivers}) is the clear
winner (see Figure~\ref{fig:2_coverage}): especially in low overhead
cases it is able to provide a much greater degree of {\bf coverage}
with respect to the other dissemination protocols (both adaptive or
not). The results obtained by \emph{Algorithm \#2: Stimuli Associated
  to Generators} are very similar to those obtained by \emph{Algorithm
  \#1}. The difference is that, in this setup the \emph{Algorithm \#1}
is unable to obtain ``overhead'' outcomes that are less than $1.37$,
this is due to its implementation and tuning. In
Table~\ref{table:adaptive_setups} are reported all the setup
parameters used to tune the algorithms considered in this part of the
performance analysis.

\begin{table}
 \caption{Adaptive protocols, parameters in different setups}
\begin{center}
	\begin{tabular}{|c||c|c|c|c|}
\hline
		Algorithm & monitoringPeriod & $\sigma$ & $\delta$ & $\alpha$ \\
\hline \hline
		\#{\bf 1} & 100 & 0.2 & 300 & $1/3$ \\
		\#{\bf 2} & 50 & 0.5 & 1000 & $3/4$ \\
		\#{\bf 3} & 50 & 0.7 & 10000 & $1$ \\
\hline
	\end{tabular}
	\label{table:adaptive_setups}
\end{center}
\end{table}

\begin{figure}
\centering
\includegraphics[width=6.0cm,angle=270]{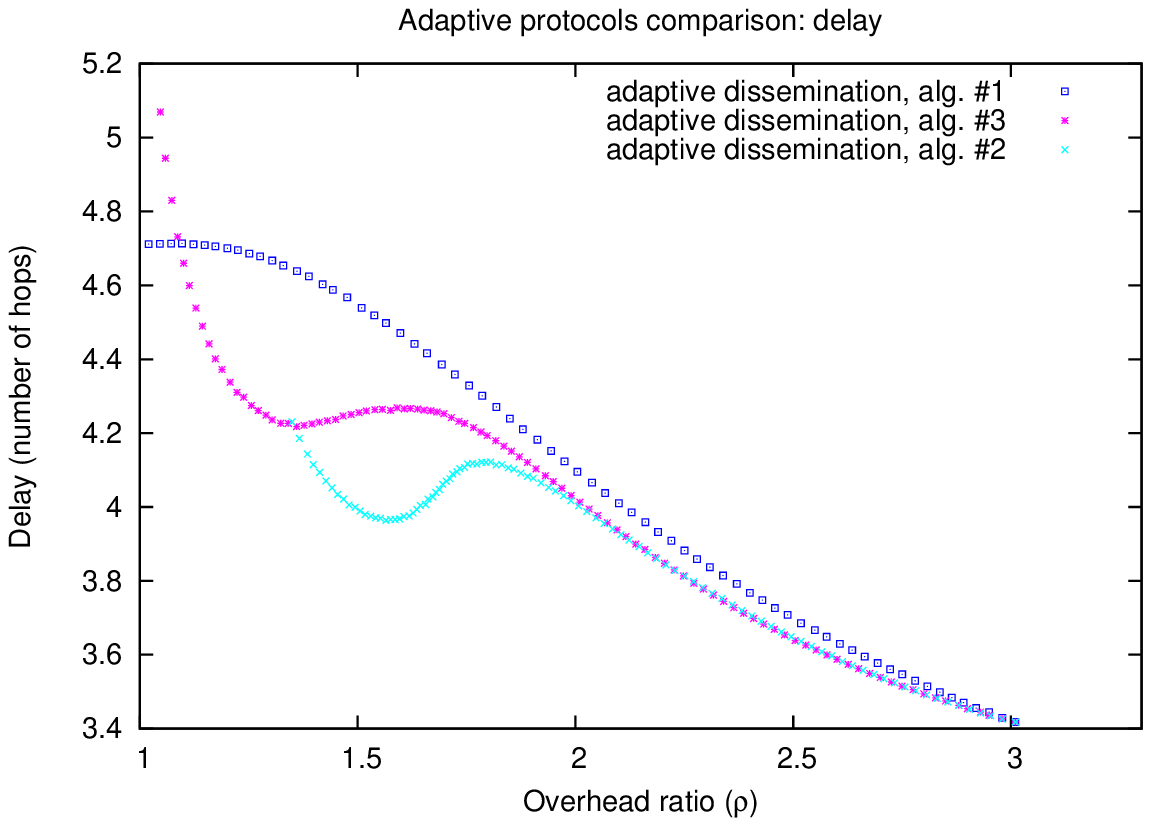}
\caption{Adaptive protocols, average delay}\label{fig:2_delay}
\end{figure}

Also under the {\bf delay} viewpoint (see Figure~\ref{fig:2_delay}),
the \emph{Algorithm \#3} is almost always better than the
counterparts. Only in case of a very low overhead (i.e.~$\rho$ that is
a little higher than $1$) the obtained delay is worse than
\emph{Algorithm \#1}. This due to the characteristics of this
dissemination algorithm: it starts with a very low dissemination
probability for each couple (generators, receivers) and only when a
reception rate that is too low is found then a probability increase is
requested. What happens, due to overhead constraint, is that this is
not sufficient to obtain a dissemination that is both efficient in
terms of coverage and fast in terms of delay. Furthermore, if $\rho$
is in the range [$1.3$, $1.7$] then the behavior of this algorithm (in
terms of {\bf delay}) is quite odd but still better than
\emph{Algorithm \#1}. The finding of a detailed and specific
motivation for this behavior is a very hard task, given the complexity
of the protocol and the many details to be considered. In the
following of this Section it will be seen that this behavior is a
characteristics of this adaptive dissemination protocol and that is
not dependent on the parameters used to setup this specific test
case. A deeper investigation is left as future work.

Finally, a few words about the \emph{Algorithm \#2}: in terms of {\bf
  coverage} it is not outstanding but in terms of {\bf delay} it shown
some interesting aspects. Also if its {\bf coverage} is very similar
to \emph{Algorithm \#1}, its {\bf delay} is much lower for almost all
$\rho$ values. In the comparison with \emph{Algorithm \#3}, its {\bf
  coverage} is a lot worse but in some parts the experienced {\bf
  delay} is quite good. This could be very interesting for user level
applications that can tolerate some packet loss but require a very
timely delivery.\\

\begin{table}
 \caption{Algorithm \#3, parameters in different setups}
\begin{center}
	\begin{tabular}{|c||c|c|c|c|}
\hline
		setup & monitoringPeriod & $\sigma$ & $\delta$ & $\alpha$ \\
\hline \hline
		\#{\bf 1} & 50 & 0.5 & 1000 & $1$ \\
		\#{\bf 2} & 50 & 0.5 & 5000 & $1$ \\
		\#{\bf 3} & 50 & 0.5 & 1000 & $3/4$ \\
		\#{\bf 4} & 50 & 0.7 & 10000 & $1$ \\
		\#{\bf 5} & 30 & 0.25 & 10000 & $1$ \\
		\#{\bf 6} & 30 & 0.25 & 10000 & $1/2$ \\
\hline
	\end{tabular}
	\label{table:alg3_setups}
\end{center}
\end{table}

In the last part of this simulation-based assessment, the focus will
be on the dissemination \emph{Algorithm \#3}: many different setups of
the algorithm will be compared. In this case, the aim is to
demonstrate that, if necessary, the protocol can be finely tuned but
in general it is pretty stable. In P2P networks, given their nature, it
is quite hard to obtain at runtime all the necessary information to
build accurate mechanisms for the fine tuning of the protocols and
algorithms that are used for dissemination or other specific
tasks. For this reason, the ``stability'' of the protocol in different
conditions is a quite interesting property.\\

\begin{figure}
\centering
\includegraphics[width=6.0cm,angle=270]{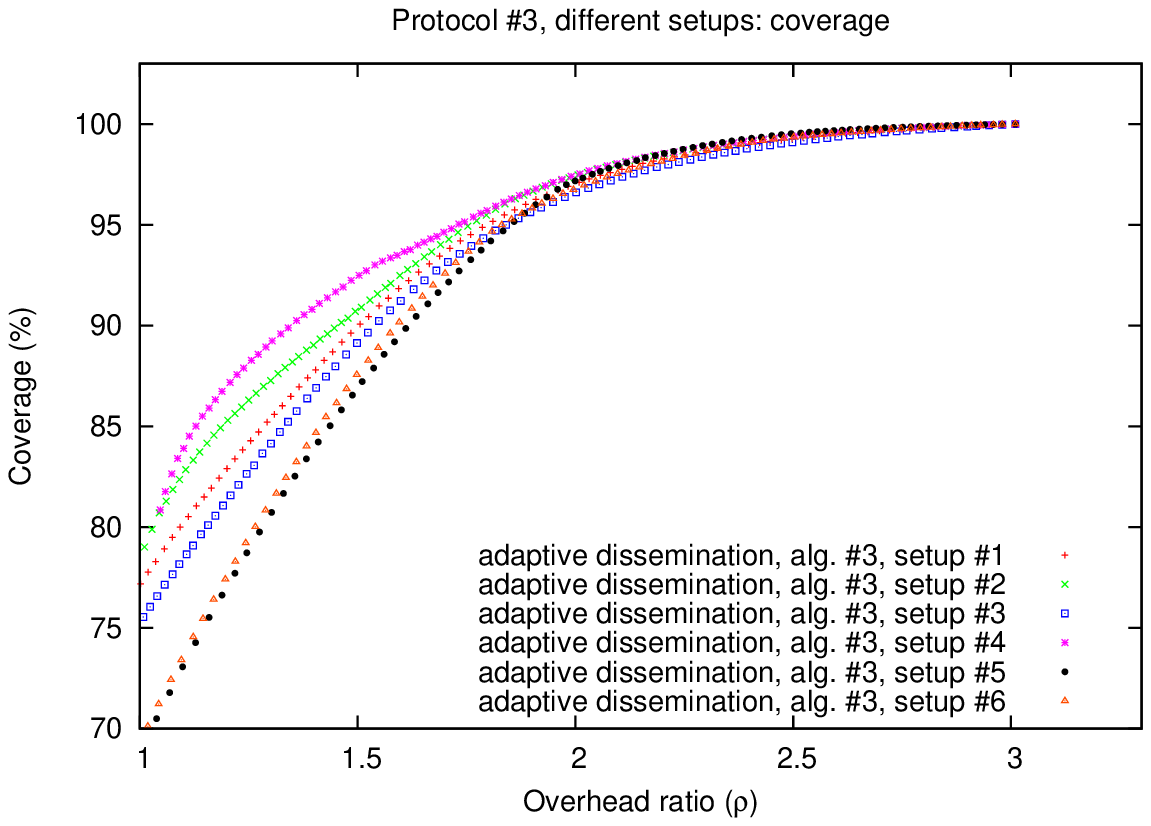}
\caption{Algorithm \#3, different setups, coverage}
\label{fig:3_coverage}
\end{figure}

\begin{figure}
\centering
\includegraphics[width=6.0cm,angle=270]{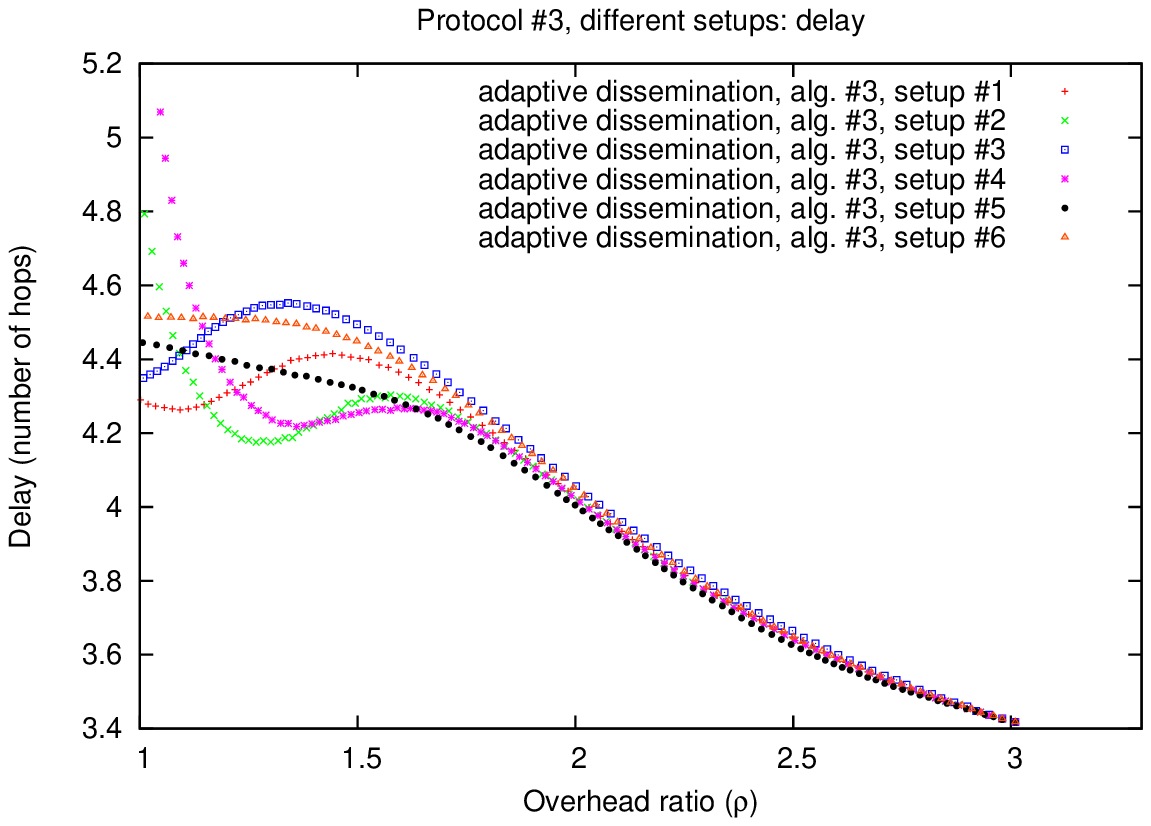}
\caption{Algorithm \#3, different setups, average delay}
\label{fig:3_delay}
\end{figure}

The parameters used for all setups are in Table~\ref{table:alg3_setups}, as
usual in Figure~\ref{fig:3_coverage} and~\ref{fig:3_delay} are shown
the average {\bf coverage} and {\bf delay} that are obtained by this
protocol with respect to the different overheads (i.e.~$\rho$
values). In Figures~\ref{fig:3_coverage} and~\ref{fig:3_delay} it can
be seen that the different setups can modify the obtained results but
always in a limited manner. In other words, the main behavior of the
adaptive algorithm is not altered in deep.

\section{Conclusions}\label{sec:conc}

In this work we described adaptive gossip protocols for data
dissemination in unstructured networks. The motivation for this work
originated in the need to efficiently disseminate events in large
Multiplayer Online Games over peer-to-peer systems. Simulation
experiments showed that adaptive gossip protocols are quite promising
in this scenario. After defining how to compare the outcomes of
different dissemination protocols and proposing some new adaptive
algorithms, we have compared their results with a couple of well-known
dissemination strategies (i.e.~\emph{Probabilistic Broadcast} and
\emph{Fixed Probability}). The results show that simple adaptive
strategies are better than non-adaptive ones both in terms of {\bf
  coverage} and {\bf delay} of data dissemination. The other main
contribution of this work is LUNES, a new freely available simulator
that is specifically aimed to the evaluation of complex protocols on
top of network graphs. LUNES is a parallel and distributed simulator
providing the necessary scalability for the evaluation of very
detailed and large-scale scenarios.\\

The obtained results are promising, and deserve further
investigations. Specifically, in this paper all the adaptive
algorithms are based on a ``positive'' stimulus. Such a stimulus
corresponds to a fixed increment of the dissemination probability,
that does not depend on the performances of the dissemination
algorithm and is activated only when they go below a certain
threshold. Probably, a more granular approach that tunes the magnitude
of the stimulus based on the performances, would permit a more fine
control of the dissemination probabilities and consequently a lower
dissemination overhead. As a further variation, we plan to implement a
more complex adaptive gossiping scheme that will be able to use both
positive and negative stimuli. In this way, each peer will be able to
fine tune the dissemination probability of each of its neighbors. We
also plan to evaluate the proposed adaptive protocols in graphs
generated with different properties (e.g.~scale-free and small-world
networks) and in presence of a larger amount of nodes. Finally, we
will investigate more in detail what is the impact of specific
parameters such as the Time-To-Live (TTL) and the cache size that so
far has not been addressed in sufficient detail.

\bibliographystyle{abbrv}
\bibliography{adaptive_gossip}

\begin{thebibliography}{10}

\bibitem{pads}
{Parallel And Distributed Simulation (PADS) Research Group}.
\newblock \url{http://pads.cs.unibo.it}, 2011.

\bibitem{Ahmed:2008}
D.~T. Ahmed and S.~Shirmohammadi.
\newblock A dynamic area of interest management and collaboration model for p2p
  mmogs.
\newblock In {\em Proceedings of the 2008 12th IEEE/ACM International Symposium
  on Distributed Simulation and Real-Time Applications}, DS-RT '08, pages
  27--34, Washington, DC, USA, 2008. IEEE Computer Society.

\bibitem{Armitage:2005}
G.~Armitage and P.~Branch.
\newblock Distribution of first person shooter online multiplayer games.
\newblock {\em Int. J. Adv. Media Commun.}, 1:59--75, September 2005.

\bibitem{Barabasi2000}
A.-L. Barab{\'a}si, R.~Albert, and H.~Jeong.
\newblock Scale-free characteristics of random networks: the topology of the
  world-wide web.
\newblock {\em Physica A: Statistical Mechanics and its Applications},
  281(1-4):69--77, Jun 2000.

\bibitem{Cronin:2004}
E.~Cronin, A.~R. Kurc, B.~Filstrup, and S.~Jamin.
\newblock An efficient synchronization mechanism for mirrored game
  architectures.
\newblock {\em Multimedia Tools Appl.}, 23:7--30, May 2004.

\bibitem{gda-ijspm-2009}
G.~D'Angelo and M.~Bracuto.
\newblock Distributed simulation of large-scale and detailed models.
\newblock {\em International Journal of Simulation and Process Modelling
  (IJSPM)}, 5(2):120--131, 2009.

\bibitem{simutools}
G.~D'Angelo and S.~Ferretti.
\newblock Simulation of scale-free networks.
\newblock In {\em Simutools '09: Proc.~of the 2nd International Conference on
  Simulation Tools and Techniques}, pages 1--10, ICST, Brussels, Belgium, 2009.
  ICST.

\bibitem{debs}
S.~Ferretti.
\newblock A synchronization protocol for supporting peer-to-peer multiplayer
  online games in overlay networks.
\newblock In {\em DEBS '08: Proc.~of the second international conference on
  Distributed event-based systems}, pages 83--94, New York, NY, USA, 2008. ACM.

\bibitem{gridpeer}
S.~Ferretti.
\newblock {Modeling Faulty, Unstructured P2P Overlays}.
\newblock In {\em Proc.~of the 19th International Conference on Computer
  Communications and Networks (ICCCN 2010)}. IEEE, August 2010.

\bibitem{disio}
S.~Ferretti and G.~D'Angelo.
\newblock Multiplayer online games over scale-free networks: a viable solution?
\newblock In {\em Proc.~of the International Workshop on DIstributed SImulation
  and Online gaming (DISIO 2010) - Conference on Simulation Tools and
  Techniques (SIMUTools 2010)}. ICST, 2010.

\bibitem{Ferretti:2006:FGH}
S.~Ferretti, C.~E. Palazzi, M.~Roccetti, G.~Pau, and M.~Gerla.
\newblock Fila in gameland, a holistic approach to a problem of many
  dimensions.
\newblock {\em Comput. Entertain.}, 4, October 2006.

\bibitem{Fletcher04unstructuredpeer-to-peer}
G.~H.~L. Fletcher and H.~A. Sheth.
\newblock Unstructured peer-to-peer networks: Topological properties and search
  performance.
\newblock In {\em 3rd International Conference on Autonomous Agents and
  MUlti-Agent Systems}, pages 14--27. Springer, 2004.

\bibitem{graphviz}
E.~R. Gansner and S.~C. North.
\newblock An open graph visualization system and its applications to software
  engineering.
\newblock {\em Softw. Pract. Exper.}, 30:1203--1233, September 2000.

\bibitem{conf/nca/GarbinatoRT07}
B.~Garbinato, D.~Rochat, and M.~Tomassini.
\newblock Impact of scale-free topologies on gossiping in ad hoc networks.
\newblock In {\em NCA}, pages 269--272. IEEE Computer Society, 2007.

\bibitem{guclu}
H.~Guclu and M.~Yuksel.
\newblock Limited scale-free overlay topologies for unstructured peer-to-peer
  networks.
\newblock {\em IEEE Trans. Parallel Distrib. Syst.}, 20(5):667--679, 2009.

\bibitem{Iimura:2004}
T.~Iimura, H.~Hazeyama, and Y.~Kadobayashi.
\newblock Zoned federation of game servers: a peer-to-peer approach to scalable
  multi-player online games.
\newblock In {\em Proceedings of 3rd ACM SIGCOMM workshop on Network and system
  support for games}, NetGames '04, pages 116--120, New York, NY, USA, 2004.
  ACM.

\bibitem{MullerFGM04}
J.~M{\"u}ller, S.~Fischer, S.~Gorlatch, and M.~Mauve.
\newblock A proxy server-network for real-time computer games.
\newblock In {\em Proceedings of Euro-Par 2004 Parallel Processing, 10th
  International Euro-Par Conference, Pisa, Italy}, volume 3149 of {\em Lecture
  Notes in Computer Science}, pages 606--613. Springer, 2004.

\bibitem{Newman03thestructure}
M.~E.~J. Newman.
\newblock The structure and function of complex networks.
\newblock {\em SIAM Review}, 45:167--256, 2003.

\bibitem{verma}
S.~Verma and W.~T. Ooi.
\newblock Controlling gossip protocol infection pattern using adaptive fanout.
\newblock In {\em ICDCS '05: Proceedings of the 25th IEEE International
  Conference on Distributed Computing Systems}, pages 665--674, Washington, DC,
  USA, 2005. IEEE Computer Society.

\bibitem{Yu:2005}
A.~P. Yu and S.~T. Vuong.
\newblock Mopar: a mobile peer-to-peer overlay architecture for interest
  management of massively multiplayer online games.
\newblock In {\em Proceedings of the international workshop on Network and
  operating systems support for digital audio and video}, NOSSDAV '05, pages
  99--104, New York, NY, USA, 2005. ACM.

\end{thebibliography}

\end{document}